\documentclass[12pt]{iopart}
\usepackage{graphicx}
\begin{document}

\title[]{Absolute versus convective helical magnetorotational instabilities
in Taylor-Couette flows}

\author{Rainer Hollerbach$^1$, Nigel Sibanda$^1$ and Eun-jin Kim$^2$}

\address{$^1$Department of Applied Mathematics,
University of Leeds, Leeds LS2 9JT, United Kingdom}
\address{$^2$School of Mathematics and Statistics,
University of Sheffield, Sheffield S3 7RH, United Kingdom}

\ead{rh@maths.leeds.ac.uk}

\begin{abstract}
We study magnetic Taylor-Couette flow in a system having
nondimensional radii $r_i=1$ and $r_o=2$, and periodic in the axial direction
with wavelengths $h\ge100$.  The rotation ratio of the inner and outer
cylinders is adjusted to be slightly in the Rayleigh-stable regime, where
magnetic fields are required to destabilize the flow, in this case triggering
the axisymmetric helical magnetorotational instability (HMRI).  Two choices of
imposed magnetic field are considered, both having the same azimuthal
component $B_\phi=r^{-1}$, but differing axial components.  The first choice
has $B_z=0.1$, and yields the familiar HMRI, consisting of unidirectionally
traveling waves.  The second choice has $B_z\approx0.1\sin(2\pi z/h)$, and
yields HMRI waves that travel in opposite directions depending on the sign of
$B_z$.  The first configuration corresponds to a convective instability, the
second to an absolute instability.  The two variants behave very similarly
regarding both linear onset as well as nonlinear equilibration.
\end{abstract}

\vspace{2pc}
\noindent{\it Keywords}: Instabilities, Pattern formation,
Magnetohydrodynamics, Taylor-Couette flows

\maketitle

\section{Introduction}

An important distinction in hydrodynamic stability theory is between absolute
and convective instabilities (Brevdo and Bridges 1996, Chomaz 2005).  This
distinction arises almost inevitably if the underlying system is such that
disturbances naturally travel in one direction only.  An absolute instability is
one where perturbations are seen to grow even if one focuses attention on a
fixed point in space, while the disturbance travels past that point.  In contrast,
a convective instability is one where perturbations grow only as seen in a
reference frame traveling with them.  In a reference frame fixed in space,
one sees only a temporary perturbation as the disturbance travels past a given
point, but a return to stability thereafter.  Absolute instability is thus a
more stringent criterion than convective instability.  The difference between
the two can be crucial, especially in bounded versus unbounded systems: in
bounded systems any global eigenmodes are necessarily absolute instabilities,
whereas in unbounded systems instabilities can arise from infinity, and the
entire concept of a global eigenmode is less clearly defined (Fox and Proctor
1998, Tobias {\it et al\/} 1998, Sandstede and Scheel 2000).

Taylor-Couette flow, the flow between differentially rotating concentric
cylinders, is one of the oldest problems in classical fluid dynamics, and
continues to attract considerable attention to this day, for both Newtonian
(Grossmann {\it et al\/} 2016) and non-Newtonian (Fardin {\it et al\/} 2014) fluids.
The basic setup by itself does not exhibit this required property that the system
should have a natural direction in which disturbances travel.  However, if an axial
through-flow is added, that does establish a preferred direction, and absolute
and convective instabilities have been widely studied in this modified problem
(e.g. Tagg {\it et al\/} 1990, Babcock {\it et al\/} 1991, Langenberg {\it et al\/}
2004, Avila {\it et al\/} 2006, Martinand {\it et al\/} 2009, Miranda-Barea {\it et
al\/} 2015).  In this work we will also consider a Taylor-Couette geometry, but the
preferred direction will be established by imposing magnetic fields, which can
break the up/down symmetry of pure, non-magnetic Taylor-Couette flow
(Knobloch 1996).

The earliest work on magnetohydrodynamic Taylor-Couette flow was by
Chandrasekhar (Chandrasekhar 1953), who showed that imposing an axial field
suppresses the onset of Taylor vortices in the standard configuration where
only the inner cylinder is rotating.  The next development was by Velikhov
(Velikhov 1959), who showed that an axial field could also have a destabilizing
effect if both cylinders are rotating, and in such a way that the system is
hydrodynamically stable according to the familiar Rayleigh (1917) criterion.
This instability, subsequently known as the magnetorotational instability (MRI),
is believed to play a key role in astrophysical accretion disks (Balbus 2003).

The MRI in a purely axial field is now also known as the `standard
magnetorotational instability' (SMRI), to distinguish it from the `helical
magnetorotational instability' (HMRI), in which a combination of axial and
azimuthal fields is imposed.  One advantage of the HMRI over the SMRI is that
the critical Reynolds number is reduced by several orders of magnitude
(Hollerbach and R\"udiger 2005), with the result that the HMRI has been obtained
experimentally (R\"udiger {\it et al\/} 2006, Stefani {\it et al\/} 2006, 2009),
whereas the standard configuration has yielded other interesting results
(Nornberg {\it et al\/} 2010, Roach {\it et al\/} 2012) but not yet the SMRI
itself.

As dramatic as the reduction in the critical Reynolds number is -- see also
Kirillov and Stefani (2010) for further insights into the scalings that cause
this -- for our purposes here the most important aspect of the HMRI is the fact
that it breaks the up/down symmetry of the system.  Imposing either purely
axial or purely azimuthal fields preserves this symmetry, but a combination of
both breaks it (Knobloch 1996).  The HMRI is thus inevitably a unidirectionally
traveling wave, and hence the above distinctions between absolute and
convective instabilities become relevant, especially since the theory was done
in a cylinder without ends, whereas the experiment necessarily had ends.

The purpose of this work therefore is to consider the HMRI in two different
fields.  First, the original combination of axial and azimuthal fields, which
amounts to a convective instability in an unbounded cylinder.  Second, a field
which is similar locally, but has an axial component that reverses periodically
on a sufficiently long axial lengthscale.  This causes waves to travel in
different directions in different portions of the cylinder, and hence yields an
absolute instability, by creating effective boundaries even in an otherwise
unbounded cylinder.  We numerically compute both linear onset and nonlinear
saturation results, and show that in both cases the absolute and convective
instability versions are still qualitatively similar.

\section{Equations}

Consider a Taylor-Couette geometry with nondimensional radii $r_i=1$ and
$r_o=2$.  Periodicity is imposed in $z$, with a wavelength $h\ge100$,
sufficiently long to give the relevant structures space to develop.  Both
cylinders rotate, at rates $\Omega_i$ and $\Omega_o$.  The values relevant to
the HMRI have $\Omega_o/\Omega_i>0.25$, just slightly into the Rayleigh-stable
regime where magnetic fields are necessary to destabilize the basic state and
trigger the onset of Taylor vortices.

In the inductionless limit appropriate to the HMRI, the Navier-Stokes and
magnetic induction equations become
$$\frac{\partial\bf U}{\partial t}=
  -\nabla p + \frac{1}{Re}\nabla^2{\bf U} - {\bf U\cdot\nabla U}
 + \frac{Ha^2}{Re}(\nabla\times{\bf b})\times{\bf B}_0,\eqno(1)$$
$$\nabla^2{\bf b}=-\nabla\times({\bf U\times B}_0),\eqno(2)$$
where $\bf U$ is the fluid flow, ${\bf B}_0$ the externally imposed
magnetic field, and $\bf b$ the induced field.  The two nondimensional
parameters are the Reynolds number
$$Re=\frac{\Omega_i r_i^2}{\nu}\eqno(3)$$
measuring the rotation rate, and the Hartmann number
$$Ha=\frac{B_0 r_i}{\sqrt{\mu\rho\nu\eta}}\eqno(4)$$
measuring the strength $B_0$ of ${\bf B}_0$.  The quantities $\mu$,
$\rho$, $\nu$ and $\eta$ are the fluid's permeability, density, viscosity and
magnetic diffusivity, respectively.

For the imposed field ${\bf B}_0$, we consider the two choices
$${\bf B}_{\rm c}=r^{-1}\,{\bf\hat e}_\phi + \delta\,{\bf\hat e}_z,\eqno(5)$$
$${\bf B}_{\rm a}=r^{-1}\,{\bf\hat e}_\phi
  + \delta\,\bigl[\sin(\kappa z)I_0(\kappa r){\bf\hat e}_z
                 -\cos(\kappa z)I_1(\kappa r){\bf\hat e}_r\bigr],\eqno(6)$$
where $\kappa=2\pi/h$, and $I_0$ and $I_1$ are the modified Bessel functions.
${\bf B}_{\rm c}$ is essentially the same as in Hollerbach and R\"udiger (2005),
except that it is more convenient here to fix the azimuthal component and have
$\delta$ as the magnitude of the axial component, rather than vice versa.  Because
${\bf B}_{\rm c}$ has no variation in $z$, the waves necessarily travel in the
same direction throughout the entire cylinder, and hence can be said to arise
`from infinity'.  The linear onsets computed by Hollerbach and R\"udiger (2005)
are thus a convective instability.

In contrast, the field ${\bf B}_{\rm a}$ has the same azimuthal component as
${\bf B}_{\rm c}$, but non-azimuthal components that vary periodically in $z$.
Using the asymptotic properties $I_0(\kappa r)\approx1$ and $I_1(\kappa r)\ll1$
for $\kappa r\ll1$, we obtain further that
$${\bf B}_{\rm a}\approx r^{-1}\,{\bf\hat e}_\phi
               + \delta\,\sin(\kappa z)\,{\bf\hat e}_z.\eqno(7)$$
The specific functions $I_0$ and $I_1$ are required only in order to satisfy
$\nabla\cdot{\bf B}_{\rm a}=0$ and $\nabla\times{\bf B}_{\rm a}= {\bf 0}$ for
the exact expression (6).  We see therefore that locally ${\bf B}_{\rm a}$ is
much the same as the $z$-independent ${\bf B}_{\rm c}$, except that its axial
component reverses sign over the global extent of the cylinders.

The presence of this $\sin(\kappa z)$ factor in the axial component of
${\bf B}_{\rm a}$ has two consequences.  First, near integer multiples of
$h/2$, where $\sin(\kappa z)\approx0$, the HMRI cannot exist at all, since it
relies crucially on both axial and azimuthal components of the imposed field
being non-zero.  Second, in between these nodes, where the axial component has
opposite signs, the waves will necessarily also travel in opposite directions,
because it is the relative sign of the axial and azimuthal components that
determines the up/down symmetry breaking (Knobloch 1996).  We see therefore
that even though the cylinders are infinite in extent, any waves that develop
cannot arise from infinity, but inevitably have to emerge at a particular
location, travel a certain distance, and then fade away again.  That is, these
waves correspond to an absolute instability, at least as far as global
eigenmodes are concerned.

These equations, together with no-slip boundary conditions for $\bf U$, and
insulating for $\bf b$, were numerically solved using an axisymmetric,
pseudo-spectral code (Hollerbach 2008).  $\bf U$ and $\bf b$ are expanded as
$${\bf U}=\nabla\times(\psi\,{\bf\hat e}_\phi) + v\,{\bf\hat e}_\phi,
\quad
  {\bf b}=\nabla\times(a\,{\bf\hat e}_\phi) + b\,{\bf\hat e}_\phi,\eqno(8)$$
then $\psi$, $v$, $a$ and $b$ are further expanded in 5400 Fourier modes in
$z$ and 50 Chebyshev polynomials in $r$.  The time-stepping of Eq.~(1) is
second-order Runge-Kutta, modified to treat the diffusive terms implicitly.
At each time-step of Eq.~(1), Eq.~(2) is inverted directly for $\bf b$.

\begin{figure}
\begin{center}
\includegraphics[width=0.65\textwidth]{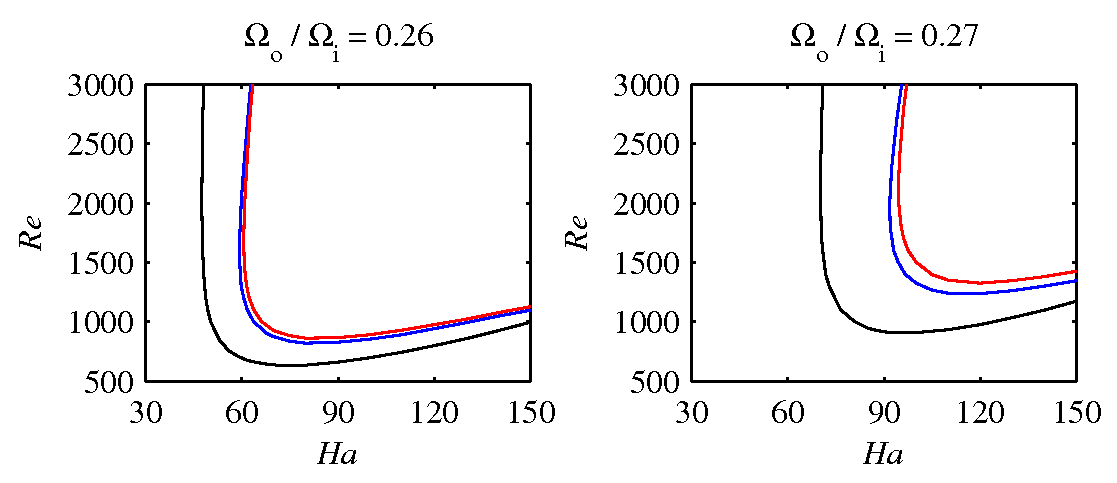}
\caption{Stability boundaries in the $(Ha,Re)$ plane, for
$\Omega_o/\Omega_i=0.26$ and $0.27$ as indicated.  In each panel the black
curves are the convective instability obtained for ${\bf B}_{\rm c}$.  The blue
and red curves are the absolute instability obtained for ${\bf B}_{\rm a}$,
with $h=100$ and $200$, respectively.  (The convective instability curves are
independent of $h$, being a one-dimensional eigenvalue problem involving
only $r$.)}
\label{fig1}
\end{center}
\end{figure}

\section{Results}

Fig.\ 1 shows how the linear onset results depend on $Ha$ and $Re$, for
$\delta=0.1$, and $\Omega_o/\Omega_i=0.26$ and $0.27$.  For parameter values
greater than these stability boundaries any infinitesimal perturbations will
grow, eventually becoming the global eigenmodes shown in later figures.  Note
how absolute instability is indeed a more stringent condition than convective
instability, requiring larger Hartmann and Reynolds numbers.  Doubling the
length of the cylinders, from $h=100$ to $200$, has almost no effect on the
absolute instability boundary; the contrast is between convective and absolute,
not the precise cylinder length in which the absolute instability is allowed to
develop.

Comparing the two rotation ratios, we see that increasing $\Omega_o/\Omega_i$
from 0.26 to 0.27 causes both instabilities to shift to larger Hartmann and
Reynolds numbers, but the shift is greater for the absolute instability
boundaries.  This is in agreement with the WKB analyses of Priede and Gerbeth
(2009) and Priede (2011), who found that the absolute instability extends less
far into the Rayleigh-stable regime than does the convective instability.

Finally, note that while the convective and absolute instability curves in
fig.\ 1 look similar, computationally they are obtained very differently.
The convective problem reduces to a one-dimensional eigenvalue formulation,
just as before in Hollerbach and R\"udiger (2005), and requires minimal
computational effort.  In contrast, the absolute problem requires a
two-dimensional time-stepping approach, and involved extensive calculations of
many different $(Ha,Re)$ combinations.

Having obtained the linear onset boundaries, we are next interested in the
spatial structure of the eigenmodes in the fully nonlinear regime.  We expect
the Taylor vortices to develop as roughly round cells in the meridional
circulation $\psi$.  Since there is relatively little structure in $r$, we
focus on $\psi$ at mid-gap only.  Simply noting whether $\psi$ there is
positive or negative is already sufficient to determine whether the
corresponding vortices are clockwise or counter-clockwise.  We can then
conveniently capture the entire space-time dynamics of the solutions in
so-called Hovm\"oller plots, showing $\psi(z,t)$ at this fixed $r=1.5$.

\begin{figure}
\begin{center}
\includegraphics[width=1\textwidth]{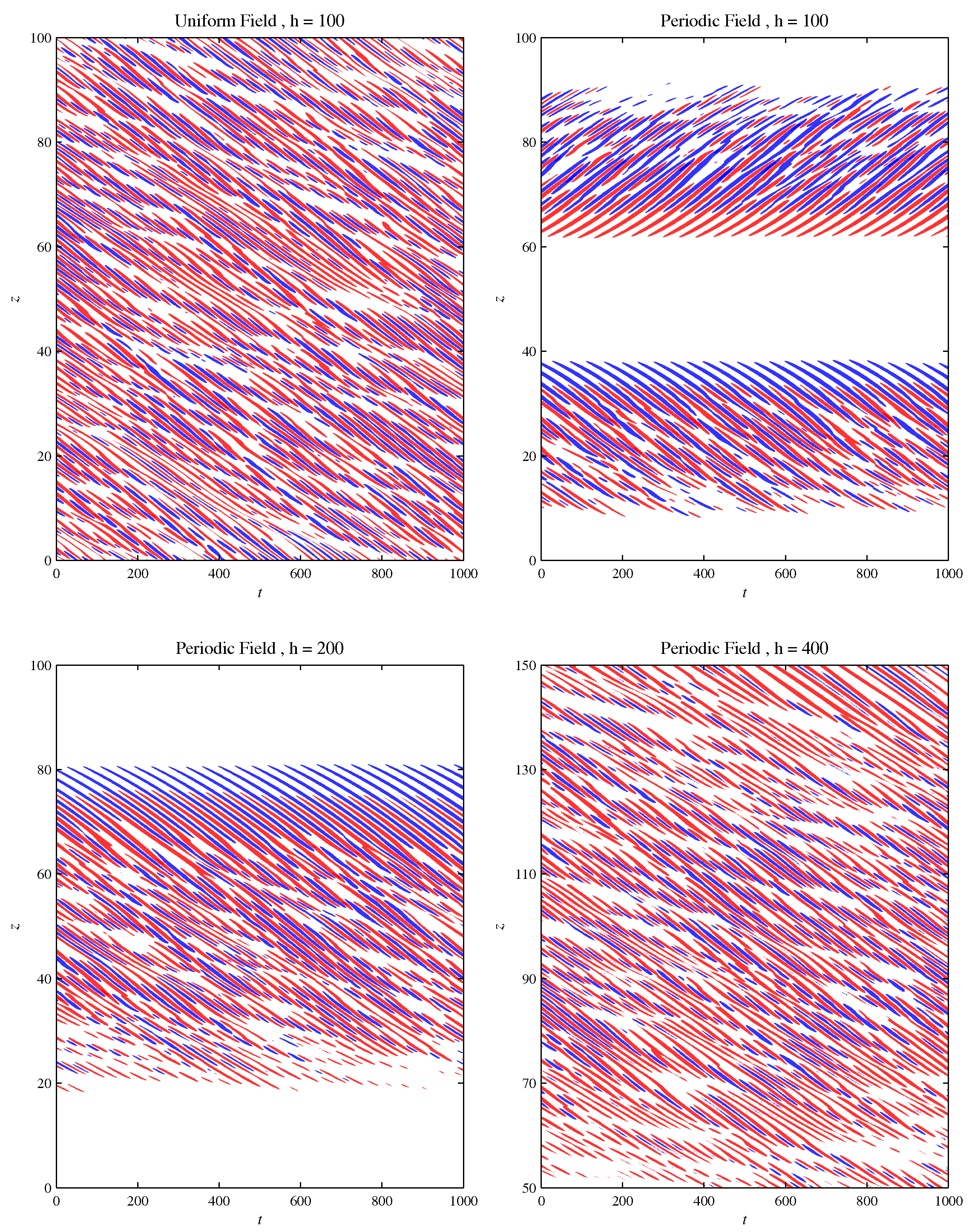}
\caption{Hovm\"oller plots showing the space-time evolution of
$\psi(z,t)$ at mid-gap, $r=1.5$.  Time is scaled on the rotational timescale
$\Omega_i^{-1}$.  Blue and red denote opposite signs, hence oppositely
rotating Taylor vortices.  ${\bf B}_{\rm c}$ with $h=100$, and
${\bf B}_{\rm a}$ with $h=100$, 200 and 400, as indicated.  For the $h=200$
and 400 results in the bottom row, only portions of the full cylinders are
shown.  All four results are for $\Omega_o/\Omega_i=0.26$, $\delta=0.1$,
$Ha=120$ and $Re=2500$.}
\label{fig2}
\end{center}
\end{figure}

Fig.\ 2 shows such plots, for ${\bf B}_{\rm c}$ with $h=100$, as
well as ${\bf B}_{\rm a}$ with $h=100$, $200$ and $400$.  For the axially
uniform field ${\bf B}_{\rm c}$, we see exactly what we would expect:
unidirectionally traveling waves that neither begin nor end at any particular
locations, but simply travel throughout the entire periodic extent of the
cylinder.  In contrast, for the axially periodic field ${\bf B}_{\rm a}$ with
$h=100$, the waves now begin in specific locations, travel in different
directions in the two halves where the axial component of ${\bf B}_{\rm a}$ is
reversed in sign, and finally die out again in specific locations.

For ${\bf B}_{\rm a}$ with $h=200$ we show only the lower half $z\in[0,100]$;
the upper half is exactly as one would expect based on the $h=100$ panel.  We
see how the waves again begin and end in particular locations.  For both the
$h=100$ and 200 results we also note how the waves are very regular and almost
periodic when they first start to grow, but gradually become more irregular and
broken-up as they travel.  This effect is stronger for $h=200$, where they have
more room to travel and develop.  Not surprisingly, it is stronger still for
$h=400$, where we show only the central portion $z\in[50,150]$.  At least in
this region, the pattern is now qualitatively essentially identical to the
original ${\bf B}_{\rm c}$ with $h=100$, even though one is a convective
instability with no beginning or ending to the waves, whereas the other is an
absolute instability with definite beginning and ending locations.

It is worthwhile also to further quantify the `starting points' of the
absolute instabilities, occurring at $z\approx38$, 80 and 165 for $h=100$, 200
and 400, respectively.  The local values of the axial field,
$\delta\sin(\kappa z)$, are then given by 0.068, 0.059 and 0.052 respectively.
For comparison, if we fix $\Omega_o/\Omega_i=0.26$, $Ha=120$ and $Re=2500$, and
instead consider $\delta$ as the control parameter, the critical value for the
onset of the \textit{convective} instability for ${\bf B}_{\rm c}$ is $\delta=
0.041$.  That is, the absolute instabilities are indeed starting in portions of
the cylinders where the field locally is convectively unstable.

The last point to note in fig.\ 2 is the existence of clearly defined phase
and group velocities for the wave patterns.  Each individual streak represents
a single Taylor vortex drifting along the cylinders, so a simple measurement of
the typical slopes yields a phase velocity of around 0.06.  Beyond that, we
see how individual streaks come and go in groups, whose edges yield a
different, smaller slope, with typical values around 0.02.  These -- fully
nonlinear -- results will be compared with linear values below.

Fig.\ 3 shows the effect of doubling the Hartmann and Reynolds numbers from
those in fig.\ 2.  The same general pattern is still observed, including even
similar phase and group velocities.  The waves are noticeably less regular
though.  The transition from almost periodic to more irregular behaviour
happens almost immediately after their formation.  Additionally, the entire
overall pattern exhibits a greater variability, with fluctuations occurring on
larger length and time scales than seen in fig.\ 2.

\begin{figure}
\begin{center}
\includegraphics[width=1\textwidth]{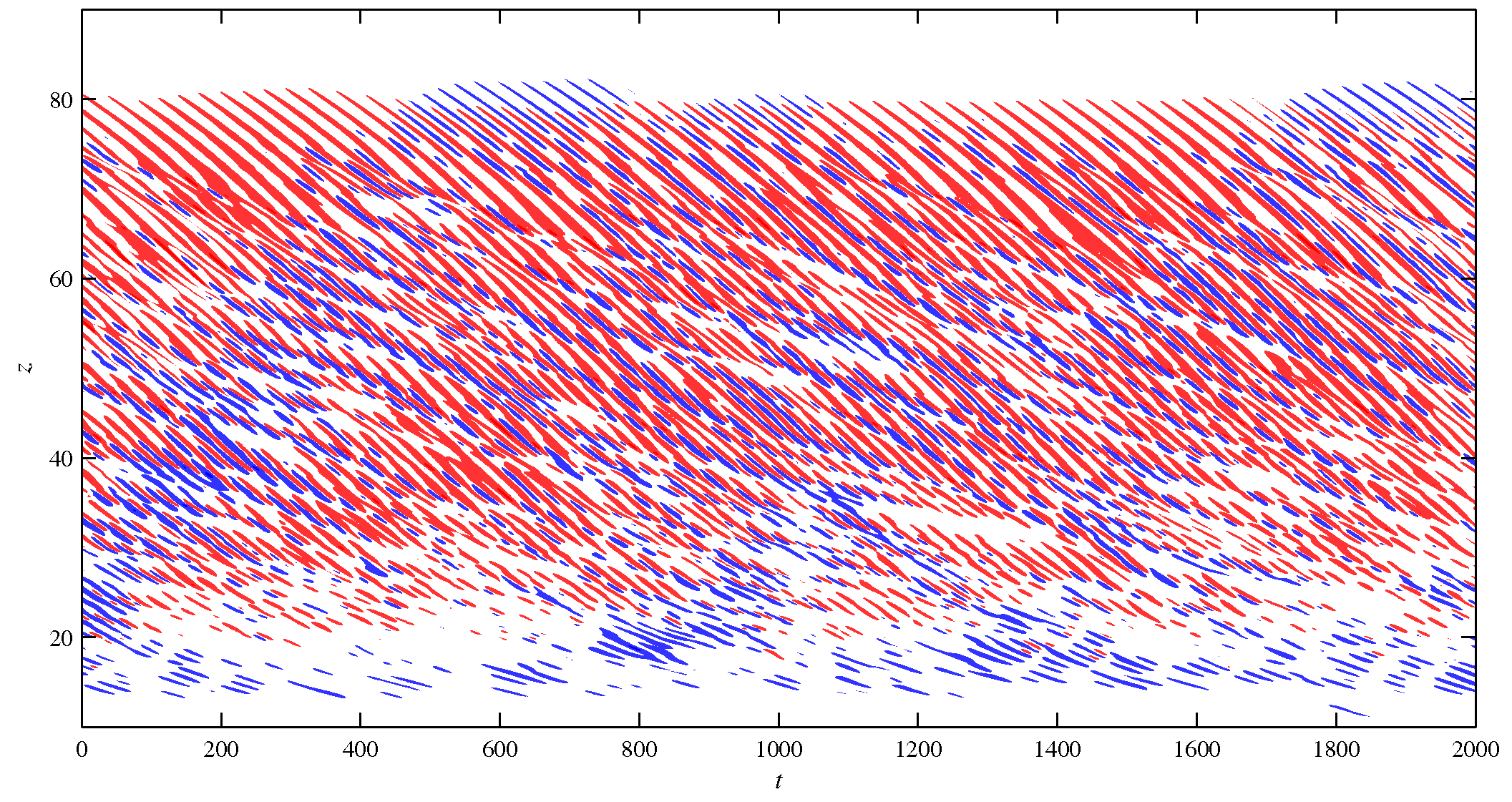}
\caption{As in fig.\ 2, but for $Ha=240$ and $Re=5000$, and
only for ${\bf B}_{\rm a}$ with $h=200$.  Note also the slightly reduced range
$z\in[10,90]$, but the longer time interval in comparison with fig.\ 2, to show
the increased temporal variability.}
\label{fig3}
\end{center}
\end{figure}

\begin{figure}
\begin{center}
\includegraphics[width=0.6\textwidth]{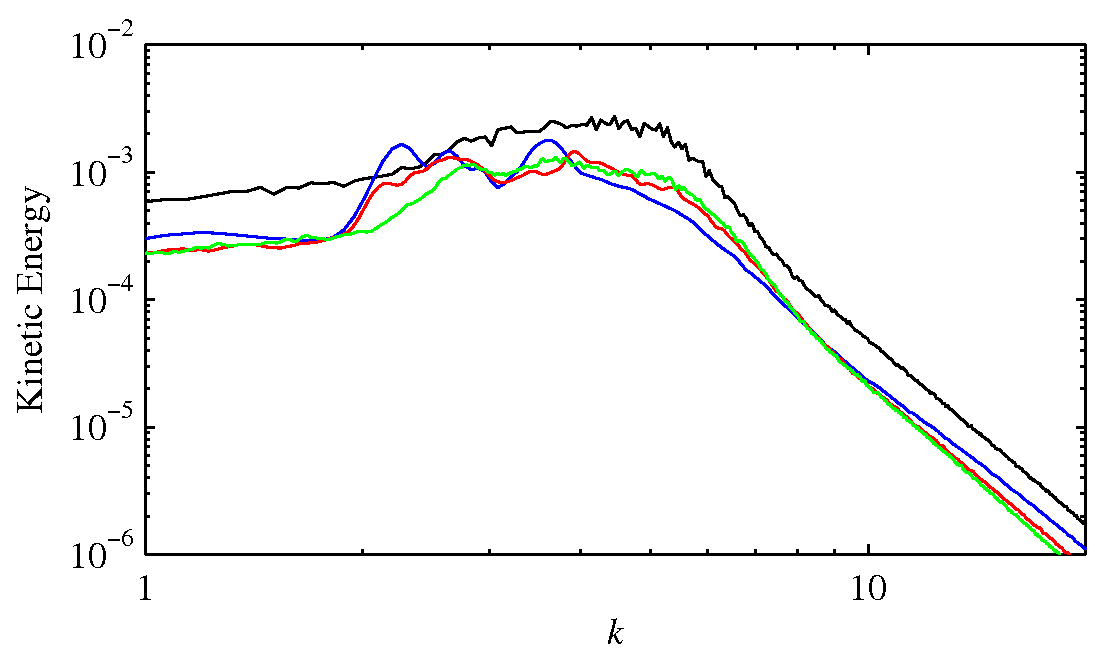}
\caption{Time-averaged axial power spectra of the solutions in
fig.\ 2.  The black line shows ${\bf B}_{\rm c}$ with $h=100$; blue, red,
green respectively denote ${\bf B}_{\rm a}$ with $h=100$, 200 and 400.  Note
also that only the range $k\in[1,20]$ is shown here, but the full computations
include both smaller $k$, corresponding to the longest lengthscales up to $h$,
and larger $k$, corresponding to dissipative lengthscales much shorter than
the typical Taylor vortex size.}
\label{fig4}
\end{center}
\end{figure}

\begin{figure}
\begin{center}
\includegraphics[width=0.6\textwidth]{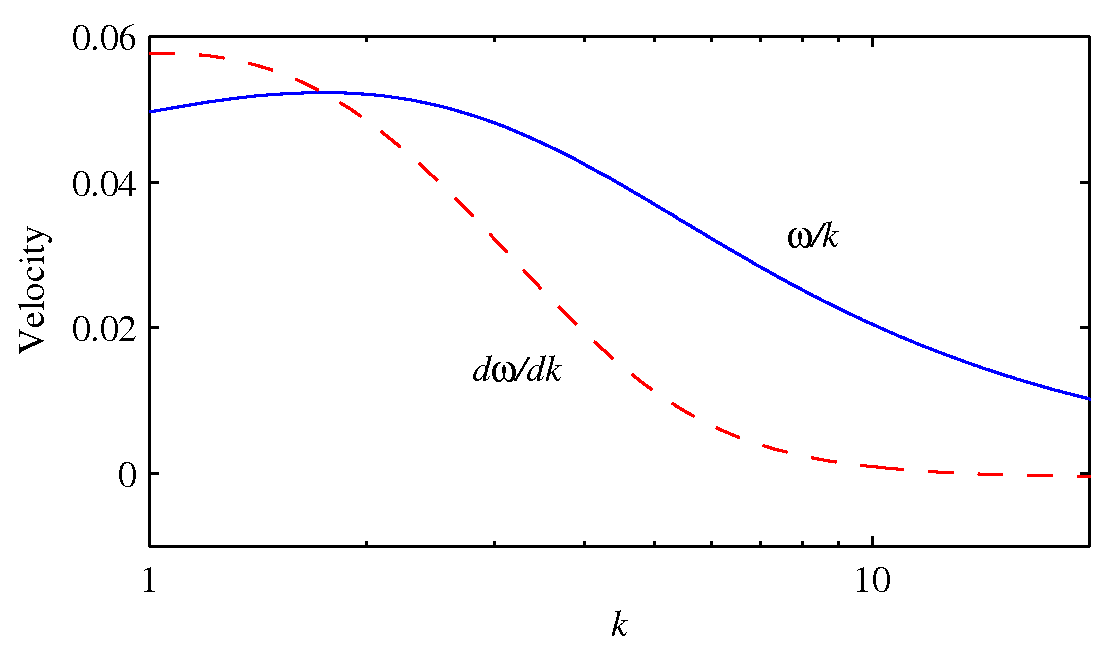}
\caption{Linear eigenvalue calculations at  $\Omega_o/\Omega_i
=0.26$, $\delta=0.1$, $Ha=120$ and $Re=2500$.  The resulting $\omega(k)$ is
then used to construct phase and group velocities $\omega/k$ and $d\omega/dk$.}
\label{fig5}
\end{center}
\end{figure}

Fig.\ 4 shows time-averaged power spectra of the solutions in fig.\ 2.  The
convective and absolute instability results are again remarkably similar, all
exhibiting a broad peak around $k\approx2-6$, indicative of the fact that
the Taylor vortices are always much the same size.  The main difference is that
the absolute results are slightly below the convective ones.  This is easily
explained by the fact that for ${\bf B}_{\rm c}$ the waves exist throughout
the entire extent of the cylinder, whereas for ${\bf B}_{\rm a}$ they only
exist in $\sim$60\% of the domain.

Finally, returning to the phase and group velocities extracted from fig.\ 2,
let us consider how well they might compare with equivalent linear results.
Fig.\ 5 shows the results of fixing $\Omega_o/\Omega_i=0.26$, $\delta=0.1$,
$Ha=120$ and $Re=2500$ in the ${\bf B}_{\rm c}$ linear eigenvalue formulation,
solving for the frequency $\omega$ (the imaginary part of the eigenvalue) as a
function of wavenumber $k$, and then plotting $\omega/k$ and $d\omega/dk$.
Both quantities vary considerably with $k$, but in the range $k\approx2-6$,
where fig.\ 4 indicates that the power is concentrated, the phase velocity
is around 0.04, and the group velocity around 0.02.  That these numbers are
close to the slopes extracted from fig.\ 2 suggests that even in the fully
nonlinear regime in figs.\ 2 and 3, there are still aspects of the solutions
that are not so different from a simple linear analysis of these waves.  See
also R\"udiger {\it et al\/} (2006), where linear wave speeds like these are
compared with the experimental results.

\section{Conclusion}

In this work we have shown how the question of convective versus absolute
instabilities can be addressed in magnetic Taylor-Couette flows even in
unbounded periodic cylinders, by imposing magnetic fields that cause the waves
to travel in different directions in different parts of the domain.  We found
that for the helical magnetorotational instability both linear and nonlinear
results are broadly similar for the convective and absolute cases, illustrating
the robust nature of the HMRI.  These results are also of interest in general
fluid dynamics and pattern formation (Hoyle 2006) contexts.

Finally, it would be worthwhile to extend this work to fully
three-dimensional solutions.  For the Hartmann numbers considered here
axisymmetric modes are strongly preferred over non-axisymmetric ones, but for
only slightly larger values even purely azimuthal fields become unstable to
non-axisymmetric instabilities, in what is now known as the azimuthal
magnetorotational instability, or AMRI (Hollerbach {\it et al\/} 2010,
Seilmayer {\it et al\/} 2014).  One might anticipate extremely complicated mode
interactions then, with the axisymmetric HMRI dominating in regions where
$|\sin(\kappa z)|\approx1$, as in this work, but the non-axisymmetric AMRI
dominating in regions where $|\sin(\kappa z)|\ll1$.  A three-dimensional,
parallelized code has recently been developed (Guseva {\it et al\/} 2015), so
further calculations in this direction are planned.

\section*{References}
\begin{harvard}

\item[]
Avila M, Meseguer A and Marques F 2006
Double Hopf bifurcation in corotating spiral Poiseuille flow
{\it Phys.\ Fluids} {\bf 18} 064101

\item[]
Babcock K L, Ahlers G and Cannell D S 1991
Noise-sustained structure in Taylor-Couette flow with through-flow
{\it Phys.\ Rev.\ Lett.} {\bf 67} 3388

\item[]
Balbus S A 2003
Enhanced angular momentum transport in accretion disks
{\it Ann.\ Rev.\ Astron.\ Astrophys.} {\bf 41} 555

\item[]
Brevdo L and Bridges T J 1996
Absolute and convective instabilities of spatially periodic flows
{\it Phil.\ Trans.\ R. Soc.\ A} {\bf 354} 1027

\item[]
Chandrasekhar S 1953
The stability of viscous flow between rotating cylinders in the presence
of a magnetic field
{\it Proc.\ R. Soc.\ A} {\bf 216} 293

\item[]
Chomaz J-M 2005
Global instabilities in spatially developing flows:
Non-normality and nonlinearity
{\it Ann.\ Rev.\ Fluid Mech.} {\bf 37} 357

\item[]
Fardin M A, Perge C and Taberlet N 2014
``The hydrogen atom of fluid dynamics'' - Introduction to
the Taylor-Couette flow for soft matter scientists
{\it Soft Matter} {\bf 10} 3523

\item[]
Fox P J and Proctor M R E 1998
Effects of distant boundaries on pattern forming instabilities
{\it Phys.\ Rev.\ E} {\bf 57} 491

\item[]
Grossmann S, Lohse D and Sun C 2016
High-Reynolds number Taylor-Couette turbulence
{\it Ann.\ Rev.\ Fluid Mech.} {\bf 48} 53

\item[]
Guseva A, Willis A P, Hollerbach R and Avila M 2015
Transition to magnetorotational turbulence in Taylor-Couette flow
with imposed azimuthal magnetic field
{\it New J. Phys.} {\bf 17} 093018

\item[]
Hollerbach R 2008
Spectral solutions of the MHD equations in cylindrical geometry
{\it Int.\ J. Pure Appl.\ Math.} {\bf 42} 575

\item[]
Hollerbach R and R\"udiger G 2005
New type of magnetorotational instability in cylindrical Taylor-Couette flow
{\it Phys.\ Rev.\ Lett.} {\bf 95} 124501

\item[]
Hollerbach R, Teeluck V and R\"udiger G 2010
Nonaxisymmetric magnetorotational instabilities in cylindrical Taylor-Couette
flow
{\it Phys.\ Rev.\ Lett.} {\bf 104} 044502

\item[]
Hoyle R B 2006 {\it Pattern Formation: An Introduction to Methods}
(Cambridge: Cambridge University Press)

\item[]
Kirillov O N and Stefani F 2010
On the relation of standard and helical magnetorotational instability
{\it Astrophys.\ J.} {\bf 712} 52

\item[]
Knobloch E 1996
Symmetry and instability in rotating hydrodynamic and magnetohydrodynamic flows
{\it Phys.\ Fluids} {\bf 8} 1446

\item[]
Langenberg J, Heise M, Pfister G and Abshagen J 2004
Convective and absolute instabilities in counter-rotating spiral Poiseuille flow
{\it Theor.\ Comp.\ Fluid Dyn.} {\bf 18} 97

\item[]
Martinand D, Serre E and Lueptow R M 2009
Absolute and convective instability of cylindrical Couette flow with axial
and radial flows
{\it Phys.\ Fluids} {\bf 21} 104102

\item[]
Miranda-Barea A, Fabrellas-Garc\'ia C, Parras L and Del Pino C 2015
Hydrodynamic instabilities in the developing region of an axially rotating
pipe flow
{\it Fluid Dyn.\ Res.} {\bf 47} 035514

\item[]
Nornberg M D, Ji H, Schartman E, Roach A and Goodman J 2010
Observation of magnetocoriolis waves in a liquid metal Taylor-Couette experiment
{\it Phys.\ Rev.\ Lett.} {\bf 104} 074501

\item[]
Priede J 2011
Inviscid helical magnetorotational instability in cylindrical Taylor-Couette flow
{\it Phys.\ Rev.\ E} {\bf 84} 066314

\item[]
Priede J and Gerbeth G 2009
Absolute versus convective helical magnetorotational instability in a
Taylor-Couette flow
{\it Phys.\ Rev.\ E} {\bf 79} 046310

\item[]
Rayleigh (Lord) 1917
On the dynamics of revolving fluids
{\it Proc.\ R. Soc.\ A} {\bf 93} 148

\item[]
Roach A H, Spence E J, Gissinger C, Edlund E M, Sloboda P, Goodman J and Ji H 2012
Observation of a free Shercliff-layer instability in cylindrical geometry
{\it Phys.\ Rev.\ Lett.} {\bf 108} 154502

\item[]
R\"udiger G, Hollerbach R, Stefani F, Gundrum T, Gerbeth G and Rosner R 2006
The traveling-wave MRI in cylindrical Taylor-Couette flow:
comparing wavelengths and speeds in theory and experiment
{\it Astrophys.\ J.} {\bf 649} L145

\item[]
Sandstede B and Scheel A 2000
Absolute and convective instabilities of waves on unbounded and large
bounded domains
{\it Physica D} {\bf 145} 233

\item[]
Seilmayer M, Galindo V, Gerbeth G, Gundrum T, Stefani F,
Gellert M, R\"udiger G, Schultz M and Hollerbach R 2014
Experimental evidence for nonaxisymmetric magnetorotational instability in a
rotating liquid metal exposed to an azimuthal magnetic field
{\it Phys.\ Rev.\ Lett.} {\bf 113} 024505

\item[]
Stefani F, Gerbeth G, Gundrum T, Hollerbach R,
Priede J, R\"udiger G and Szklarski J 2009
Helical magnetorotational instability in a Taylor-Couette flow with strongly
reduced Ekman pumping
{\it Phys.\ Rev.\ E} {\bf 80} 066303

\item[]
Stefani F, Gundrum T, Gerbeth G, R\"udiger G,
Schultz M, Szklarski J and Hollerbach R 2006
Experimental evidence for magnetorotational instability in a Taylor-Couette
flow under the influence of a helical magnetic field
{\it Phys.\ Rev.\ Lett.} {\bf 97} 184502

\item[]
Tagg R, Edwards W S and Swinney H L 1990
Convective versus absolute instability in flow between counterrotating cylinders
{\it Phys.\ Rev.\ A} {\bf 42} 831

\item[]
Tobias S M, Proctor M R E and Knobloch E 1998
Convective and absolute instabilities of fluid flows in finite geometry
{\it Physica D} {\bf 113} 43

\item[]
Velikhov E P 1959
Stability of an ideally conducting liquid flowing between cylinders rotating
in a magnetic field
{\it Sov.\ Phys.\ JETP} {\bf 36} 995

\end{harvard}

\end{document}